\documentclass[fleqn]{article}
\usepackage{amsmath,amssymb}
\usepackage{epsfig,graphicx}

\catcode`@=11
\catcode`@=12

\begin{document}
\title{\vskip-1.7cm \bf Dark matter as a ghost free conformal extension of Einstein theory}
\date{}
\author{A.O.Barvinsky}
\maketitle
\hspace{-8mm} {\,\,\em Theory Department, Lebedev
Physics Institute, Leninsky Prospect 53, Moscow 119991, Russia}

\begin{abstract}
We discuss ghost free models of the recently suggested mimetic dark matter theory. This theory is shown to be a conformal extension of Einstein general relativity. Dark matter originates from gauging out its local Weyl invariance as an extra degree of freedom which describes a potential flow of the pressureless perfect fluid. For a positive energy density of this fluid the theory is free of ghost instabilities, which gives strong preference to stable configurations with a positive scalar curvature and trace of the matter stress tensor. Instabilities caused by caustics of the geodesic flow, inherent in this model, serve as a motivation for an alternative conformal extension of Einstein theory, based on the generalized Proca vector field. A potential part of this field modifies the inflationary stage in cosmology, whereas its rotational part at the post inflationary epoch might simulate rotating flows of dark matter.
\end{abstract}

\maketitle

\section{Introduction}

Recently suggested model of mimetic dark matter \cite{ChamsMukh} is based on the GR action of the gravitational $g_{\mu\nu}^{\rm phys}$ and matter $\varphi$ fields
    \begin{eqnarray}
    &&S[\,g_{\mu\nu}^{\rm phys},\varphi\,]
    =\int d^4x\,g^{1/2}_{\rm phys}
    \left(\,\frac12\, R(\,g_{\mu\nu}^{\rm phys})
    +L(g_{\mu\nu}^{\rm phys},
    \varphi,\partial\varphi) \right), \label{action00}
    \end{eqnarray}
in which the physical metric $g_{\mu\nu}^{\rm phys}$ is parameterized in terms of the fundamental metric $g_{\mu\nu}$ and the spacetime gradients of the scalar field $\phi$.
    \begin{eqnarray}
    g_{\mu\nu}^{\rm phys}=-(g^{\alpha\beta}
    \partial_\alpha\phi
    \partial_\beta\phi)\,g_{\mu\nu}
    \equiv \varPhi^2 g_{\mu\nu}            \label{change}
    \end{eqnarray}
This action in terms of new variables $S[\,g_{\mu\nu}^{\rm phys}(g_{\mu\nu},\phi),\varphi\,]$ generates variational equations with respect to $g_{\mu\nu}$ and $\phi$
    \begin{eqnarray}
    &&G^{\mu\nu}_{\rm phys}
    =T^{\mu\nu}_{\rm phys}
    +\varepsilon\,u^\mu u^\nu,  \label{Eq1}\\
    &&\nabla_\mu^{\rm phys}
    (\varepsilon u^\mu)=0,      \label{continuity}\\
    &&\varepsilon=
    R_{\rm phys}+T_{\rm phys}.   \label{density}
    \end{eqnarray}
where the covariant derivative $\nabla_\mu^{\rm phys}$, the Einstein tensor $G^{\mu\nu}_{\rm phys}$, the Ricci scalar $R_{\rm phys}$ and matter stress tensor
    \begin{eqnarray}
    T^{\mu\nu}_{\rm phys}=\frac2{g^{1/2}_{\rm phys}}\,
    \frac{\delta}{\delta g_{\mu\nu}^{\rm phys}}
    \int d^4x\,g^{1/2}_{\rm phys} L(g_{\mu\nu}^{\rm phys},
    \varphi,\partial\varphi)
    \end{eqnarray}
are determined with respect to the physical metric, as well as their traces, $T_{\rm phys}=g_{\mu\nu}^{\rm phys}T^{\mu\nu}_{\rm phys}$, etc. The vector $u_\mu$ is a four-velocity generated by the velocity potential $\phi$ (we consider the case of a timelike $u_\mu$ and work in the $(-+++)$ metric signature),
    \begin{eqnarray}
    u_\mu=\partial_\mu\phi,\quad
    g^{\mu\nu}_{\rm phys}u_\mu u_\nu=-1. \label{kinematical}
    \end{eqnarray}
Note that this normalization to unity in the {\em physical} metric is a kinematical relation -- the corollary of (\ref{change}) independent of dynamics.

Eqs.(\ref{Eq1})-(\ref{density}) differ from those of the original action (\ref{action00}) by an extra ``matter" source -- pressureless dust fluid with four-velocity $u_\mu$ and density (\ref{density}) satisfying the continuity equation (\ref{continuity}). As it was suggested in \cite{ChamsMukh} this dust can play the role of dark matter, whose imprint on large scale structure of the Universe can survive till now provided one includes a proper coupling of the scalar $\phi$ to the inflaton $\varphi$ in the matter Lagrangian.

The explanation of the paradox that a simple reparametrization of variables (\ref{change}) can lead to extra new solutions of equations of motion which differ from those of the original GR equations $G_{\mu\nu}=T_{\mu\nu}$ is as follows. Point is that the change of variables from the original ten components of $g_{\mu\nu}^{\rm phys}$ to ten new metric coefficients $g_{\mu\nu}$ is not invertible even for fixed $\phi$.\footnote{For free $\phi$ the transformation $g_{\mu\nu}^{\rm phys}\to (g_{\mu\nu},\phi)$ is of course not one to one, because this is a map from ten variables to eleven.} The original physical metric in terms of the fundamental metric $g_{\mu\nu}$ is conformally invariant, so that the theory in terms of new variables has local Weyl invariance with respect to the transformation of the metric
    \begin{eqnarray}
    \Delta_\sigma g_{\mu\nu}(x)=\sigma(x)\,g_{\mu\nu}(x), \quad
    \Delta_\sigma S[\,g_{\mu\nu}^{\rm phys}(g_{\mu\nu},\phi),\varphi\,]=0 \label{conftrans}
    \end{eqnarray}
with an arbitrary function $\sigma(x)$. Therefore, it generates identically traceless Eq.(\ref{Eq1}) and requires the procedure of conformal gauge fixing. A natural conformal gauge can be chosen as
    \begin{eqnarray}
    g^{\mu\nu}
    \partial_\mu\phi\,
    \partial_\nu\phi+1=0.            \label{gauge0}
    \end{eqnarray}
Its advantage is that it immediately allows one to identify the fundamental metric with the physical one $g_{\mu\nu}=g_{\mu\nu}^{\rm phys}$ and remove label ``phys" in all the equations (\ref{Eq1})-(\ref{density}). This, actually, gives a reinterpretation of the kinematical relation in (\ref{kinematical}), which becomes now a gauge condition in the local gauge-invariant theory with the action $S[\,g_{\mu\nu}^{\rm phys}(g_{\mu\nu},\phi),\varphi\,]$.

Thus, the model of \cite{ChamsMukh} turns out to be a conformal extension of Einstein theory, which is local Weyl invariant in terms of the {\em fundamental} metric field $g_{\mu\nu}$. Similar extensions of general relativity were repeatedly used for various purposes, including the attempts of avoiding conformal anomalies \cite{FradkinVilkovisky} or embedding the Einstein theory into Weyl invariant gravity \cite{Maldacena}. However, in contrast to the conformal {\em off-shell} extension suggested in \cite{FradkinVilkovisky}, which preserves Einstein theory on-shell and only modifies its off-shell effective action, here the Einstein theory is modified already at the classical level and acquires extra degree of freedom of a pressureless perfect fluid. According to \cite{ChamsMukh} this fluid can mimic the behavior of a real cold dark matter.

Primary check on quantum consistency of this model is its stability with respect to possible ghost modes. This issue was not exhaustively considered in \cite{ChamsMukh}. So here we show by explicitly calculating quadratic part of the action on the generic background that the theory is free of ghosts whenever this background satisfies positive energy condition $\varepsilon>0$. For this reason we develop the Lagrangian and canonical formalism of this theory in the gauge (\ref{gauge0}) in which the latter emerges as one of the equations of motion. Then we show that the dynamical degree of freedom of the dark matter fluid is free of ghosts, though it can still suffer from caustic instabilities. Finally we suggest a Proca vector field which can model also rotational flows of dark matter (which are not available in the model of \cite{ChamsMukh}). Proca nature of the vector field guarantees it from ghost instabilities. For non-rotational configurations this model implies algebraic relation between the dark matter density and the inflaton field, so that they both decay in the end of inflation and cannot simulate real cold dark matter. However, the rotational part of the vector field might mimic real dark matter and its adiabatic perturbations.

\section{Mimetic dark matter and gauged out Weyl invariance}
Local conformal invariance of the theory (\ref{action00}) implies a gauge fixing procedure which includes imposing a relevant conformal gauge $\chi(g_{\mu\nu},\phi)=0$ and adding, under quantization, the Faddeev-Popov ghost determinant ${\rm Det}\,Q$ into the path integral measure (here $Q$ is the Faddeev-Popov operator which determines the transformation of the gauge condition, $\Delta_\sigma\chi=Q\sigma$). Choosing as $\chi(g_{\mu\nu},\phi)$ the left hand side of (\ref{gauge0}) and representing its delta-function as the integral over the Lagrange multiplier $\varepsilon$ we get
    \begin{eqnarray}
    &&Z=\int D[\,g_{\mu\nu},\phi\,]\,
    e^{iS[\,g_{\mu\nu}^{\rm phys}(g_{\mu\nu},\phi)\,]}\,{\rm Det}\,Q\,
    \delta[\,g^{\mu\nu}
    \partial_\mu\phi\,\partial_\nu\phi+1\,]
    =\int D[\,g_{\mu\nu},\phi\,]\,D\varepsilon\,
    e^{iS[\,g_{\mu\nu},\phi,\varepsilon\,]},   \label{Z}\\
    &&S[\,g_{\mu\nu},\phi,\varepsilon\,]
    =\int d^4x\,g^{1/2}
    \left(\,\frac12\,R-\frac12\, \varepsilon\,(g^{\mu\nu}\partial_\mu\phi\,
    \partial_\nu\phi+1)
    +L \right).               \label{action}
    \end{eqnarray}
Here for brevity we omitted integration over matter fields $\varphi$ and introduced the gauge fixed version of the action (\ref{action00}) enforcing the conformal gauge via the Lagrange multiplier $\varepsilon$. Note that in view of the delta-function type gauge in the integrand of (\ref{Z}) the argument $g_{\mu\nu}^{\rm phys}(g_{\mu\nu},\phi)$ of the original action can be replaced by the fundamental field $g_{\mu\nu}$.

Note also that the Faddeev-Popov gauge fixing procedure for diffeomorphism invariance is implicit in the canonical integration measure $D[\,g_{\mu\nu},\phi\,]$. As far as it is concerned the conformal invariance, its ghost determinant ${\rm Det}\,Q$ is trivial because for the conformal gauge (\ref{gauge0}) the Weyl transformation (\ref{conftrans}) yields a unit operator,
    \begin{eqnarray}
    \Delta_\sigma[\,g^{\alpha\beta}
    \partial_\alpha\phi\,\partial_\beta\phi+1\,]=
    -\sigma\,g^{\alpha\beta}
    \partial_\alpha\phi\,\partial_\beta\phi=\sigma.
    \end{eqnarray}
Thus, the gauged out theory is described by the action (\ref{action}) where the matter Lagrangian $L=L(g_{\mu\nu},\varphi,\partial\varphi,\phi)$ may include arbitrary matter fields $\varphi$ and their interaction with the scalar $\phi$. All its classical and quantum effects are described by the path integral generating functional (\ref{Z}) (with additional path integration over matter fields $\varphi$). In what follows we will consider this theory in the tree-level approximation and analyze its quantum consistency with regard to ghost instability.

Variations with respect to $\varepsilon$, $\phi$ and $g_{\mu\nu}$ give respectively
    \begin{eqnarray}
    &&\phi_\mu^2=-1, \quad
    \phi_\mu\equiv\partial_\mu\phi,     \label{1}\\
    &&\nabla_\mu
    \big(\varepsilon\phi^\mu\big)=0,        \label{2}\\
    &&G^{\mu\nu}-\varepsilon\,
    \phi^\mu\phi^\nu-T^{\mu\nu}=0
    \end{eqnarray}
and the trace of the last equation gives
    \begin{eqnarray}
    \varepsilon=R+T,
    \end{eqnarray}
so that the system of equations, as it is expected, becomes equivalent to (\ref{Eq1})-(\ref{density}) with $g_{\mu\nu}^{\rm phys}=g_{\mu\nu}$. Thus, indeed mimetic dark matter arises as a conformal extension of the Einstein theory, its density playing the role of the Lagrange multiplier for gauged out Weyl invariance.

The form of the action (\ref{action}) clearly shows that the theory does not have higher-derivative ghosts. They could have been expected to arise in (\ref{action00}) under the conformal transformation (\ref{change}), $g_{\mu\nu}^{\rm phys}=\varPhi^2g_{\mu\nu}$, in view the well-known relation \cite{FradkinVilkovisky}
    \begin{eqnarray}
    \int d^4x\,g^{1/2}_{\rm phys}
    R(\,g_{\mu\nu}^{\rm phys})=
    \int d^4x\,g^{1/2}\Big(R(\,g_{\mu\nu})\,\varPhi^2+
    6\,g^{\mu\nu}\partial_\mu\varPhi\,\partial_\nu\varPhi\Big),
    \quad \varPhi=\sqrt{-g^{\alpha\beta}
    \partial_\alpha\phi\,\partial_\beta\phi}.
    \end{eqnarray}
The second term on the right hand side here generates fourth-order derivatives in equations of motion, but this term can be identically canceled by the conformal gauge breaking term $-6[\nabla_\mu(\varPhi-1)]^2=-6(\nabla_\mu\varPhi)^2$ instead of the delta-function type gauge in (\ref{Z}). This leaves us with the gauged out Lagrangian of the second order in derivatives, $g^{1/2}(-g^{\alpha\beta} \partial_\alpha\phi\,\partial_\beta\phi)\,R(\,g_{\mu\nu})$, which as one can easily show again leads to mimetic DM equations. However, it is not yet guaranteed that the extra degree of freedom comprised by the fields $\phi$ and $\varepsilon$ is free of ghost instabilities. The domain of their ghost stability is considered in the next section.

\section{Absence of ghosts}

Dynamical properties of the field $\phi$ and the Lagrange multiplier $\varepsilon$ follow from their canonical formalism. The latter is easily available on a flat-space background, which is sufficient for the analysis of kinetic terms of all degrees of freedom \cite{BlaPuSib}. We have the canonical momentum conjugated to $\phi$
    \begin{eqnarray}
    p=\varepsilon \dot\phi,  \label{p-sigma}
    \end{eqnarray}
and the Hamiltonian
    \begin{eqnarray}
    H=\frac12\,\frac{p^2}{\varepsilon}
    +\frac12\,\varepsilon\,(1+\phi_i^2),
    \quad \phi_i\equiv\partial_i\phi.
    \end{eqnarray}

Now the Lagrangian multiplier can be excluded by the equation $\partial H/\partial\varepsilon=0$ (contrary to the Lagrangian formalism where it was impossible in view of linearity of (\ref{action}) in $\varepsilon$),
    \begin{eqnarray}
    \varepsilon=\frac{p}{\sqrt{1+\phi_i^2}}.\label{varepsilon}
    \end{eqnarray}
With this $\varepsilon$ the Hamiltonian becomes {\em linear in momentum}, $H=p\sqrt{1+\phi_i^2}$, and the canonical action becomes \cite{BlaPuSib}
    \begin{eqnarray}
    S=\int dt\,d^3x\,\left(p\,\dot\phi-p\,
    \sqrt{1+\phi_i^2}\right).               \label{canonical}
    \end{eqnarray}
Opposite sign of the square root in (\ref{varepsilon}) leads to flipping the sign of the Hamiltonian, but the resulting action remains equivalent to (\ref{canonical}) because of the trivial canonical transformation $(\phi,p)\to -(\phi,p)$.

Equations of motion for phase space variables $(\phi,p)$ read
    \begin{eqnarray}
    &&\dot\phi=\sqrt{1+\phi_i^2}, \label{phiequation}\\
    &&\dot p=
    \partial_i\left(\frac{p\,\phi^i}
    {\sqrt{1+\phi_i^2}}\right).     \label{pequation}
    \end{eqnarray}
The first one is obviously the norm of the 4-velocity (\ref{1}), whereas the second is just the continuity equation (\ref{2}) with $p=\varepsilon\dot\phi$.

If we repeat all these steps in curved spacetime, the final result for the canonical form of the $(\phi,\varepsilon)$-part of the action (\ref{action}) is
    \begin{eqnarray}
    S=\int d^4x\,\left(p\,\dot\phi
    -N\,p\,\sqrt{1+g^{ij}\partial_i\phi\partial_j\phi}
    -N^ip\,\partial_i\phi\right),
    \end{eqnarray}
where $(N,N^i)$ are the lapse and shift functions, and their coefficients are respectively the canonical superhamiltonian and supermomenta of the scalar field.

Ghost modes arise in the theory when its kinetic term in the Lagrangian action is not positive definite and can dynamically evolve to $-\infty$ for large time derivatives of the field. An obvious difficulty with (\ref{canonical}) is that the canonical momentum $p$ cannot be excluded from the action by its variational equation. However, the Lagrangian form of the action can be obtained by switching the roles of momenta and coordinates. The configuration space coordinate $\phi$ can in principle be expressed from Eq.(\ref{pequation}) in terms of $p$ and $\dot p$, and the sign of the kinetic term as a function of $\dot p$ would indicate the nature of this mode. Exact solution of (\ref{pequation}) for $\phi$ is not available, so that we shall have to analyze the situation in the linearized theory on the generic background.

Decomposing the dust variables $\phi\to\phi+\varphi$, $p\to p+\pi$, into their background values and perturbations $(\varphi,\pi)$ we have the quadratic part of the canonical action
    \begin{eqnarray}
    S_{(2)}=\int d^4x\,\left\{-\dot\pi\,\varphi
    -\frac12\,\varepsilon\,
    (\delta^{ik}-v^i v^k)\,\partial_i\varphi\,\partial_k\varphi
    -v^i\partial_i\varphi\,\pi\right\},   \label{S_2}
    \end{eqnarray}
where we integrated by parts the symplectic term $\pi\dot\varphi$ with respect to time and introduced a notation $v^i$ for the 3-dimensional velocity of the background dust and recovered a background value of $\varepsilon$, cf. Eq.(\ref{p-sigma}),
    \begin{eqnarray}
    &&v^i\equiv\frac{\phi^i} {\sqrt{1+\phi_i^2}},\\
    &&\varepsilon=\frac{p}{\dot\phi}=\frac1{\sqrt{1+\phi_i^2}}.
    \end{eqnarray}
In terms of it the linearized equation (\ref{pequation}) reads
    \begin{eqnarray}
    &&\dot\pi-\partial_i(\pi v^i)=
    \tilde\Delta\varphi,\\
    &&\tilde\Delta\equiv\partial_i\big[\varepsilon(\delta^{ik}-v^i v^k)\,\partial_k\big].
    \end{eqnarray}
Here, in view of the fact that
    \begin{eqnarray}
    v^2\equiv v_i^2=
    \frac{\phi_i^2}{1+\phi_i^2}<1,
    \end{eqnarray}
the generalized Laplacian $\tilde\Delta$ involves squares of all spatial derivatives both parallel to the velocity vector, $(\,||\,)$, and transversal to it, $(\,\perp\,)$,
    \begin{eqnarray}
    \tilde\Delta=\varepsilon\Delta_\perp+\varepsilon(1-v^2)\partial_\|^2
    +O(\partial^1),\quad
    \Delta_\perp=\left(\delta^{ik}-\frac{v^i v^k}{v^2}\right)\partial_i\partial_k
    \end{eqnarray}
(only in the ultrarelativistic limit $v^2\to 1$ it tends to the 2-dimensional Laplacian $\Delta_\perp$ acting in the plane transversal to $v^i$). Under appropriate boundary conditions it is invertible and gives the nonlocal in space expression for $\varphi$ in terms of $\pi$ and $\dot\pi$
    \begin{eqnarray}
    &&\varphi=\frac1{\tilde\Delta}\,\big(\dot\pi
    -\partial_i(\pi v^i)\big).
    \end{eqnarray}
Substituting it into (\ref{S_2}) we obtain the Lagrangian action in terms of $\pi$ and $\dot\pi$, which is nonlocal in space, but local in time,
    \begin{eqnarray}
    S_{(2)}=-\int d^4x\,\big(\dot\pi
    -\partial_i(\pi v^i)\big)\frac1{\tilde\Delta}\,\big(\dot\pi
    -\partial_k(\pi v^k)\big).
    \end{eqnarray}
Under zero boundary conditions at infinity and {\em positive} $\varepsilon$ the operator $\tilde\Delta$ is negative definite (and for short wavelengths modes it is negative definite independently of boundary conditions at infrared infinity). Therefore quadratic in momentum ``velocities" $\dot\pi$ part of the Lagrangian is positive, which implies that the theory is free of ghost instabilities for $\varepsilon>0$.

Physically this is a very natural criterion which coincides with the positive energy condition for a dust fluid with the stress tensor $T_{\mu\nu}=\varepsilon\,u_\mu u_\nu$, $\varepsilon>0$. Since $\varepsilon$ is expressed here in terms of scalar curvature and matter stress tensor, the ghost stability imposes the bound
    \begin{eqnarray}
    R+T>0
    \end{eqnarray}
and gives strong preference to dS-type backgrounds with a positive cosmological constant.

Another type of instability which is perhaps not so dangerous at the quantum level is due to formation of caustics. They are inevitable for generic geodesic flow which is associated with the potential $\phi$ satisfying the Hamilton-Jacobi equation (\ref{gauge0}). Field ``dust" moves along geodesics -- the characteristic curves of this equation -- and forms caustic singularities in view of its pressureless nature (see discussion of this phenomenon in context of field models of dark matter \cite{caustics} and also Horava and ghost condensation gravity models \cite{Mukohyama,Mukohyamaetal,BlaPuSib}). Eq.(\ref{gauge0}) and its geodesics are artifacts of conformal gauge fixing in our approach. However, this particular gauge has a distinguished status because in this gauge the physical conformally invariant metric $g^{\rm phys}_{\mu\nu}$ coincides with the auxiliary metric $g_{\mu\nu}$ (or in the original formulation of \cite{ChamsMukh} this is a kinematical relation (\ref{kinematical})). Therefore, this problem cannot be circumvented by an alternative conformal gauge fixing and remains a serious difficulty.

\section{Vector field model of DM}

In addition to the caustic problem the model of \cite{ChamsMukh} does not admit rotating dark matter because of the potential flow of the 4-velocity $u_\mu=\partial_\mu\phi$. These limitations might perhaps be circumvented within the vector field model with the physical metric parameterized by the dynamical vector field $u_\mu$
    \begin{eqnarray}
    g_{\mu\nu}^{\rm phys}=-(g^{\alpha\beta}
    u_\alpha u_\beta)\,g_{\mu\nu}.
    \end{eqnarray}
We may start with the action in terms of the physical metric which also contains the Maxwell kinetic term to make this vector field propagating
    \begin{eqnarray}
    &&S[\,g_{\mu\nu}^{\rm phys},\varphi\,]
    =\int d^4x\,g^{1/2}_{\rm phys}
    \left(\,\frac12\,R(\,g_{\mu\nu}^{\rm phys})
    +L(g_{\mu\nu}^{\rm phys},\varphi,\partial\varphi,u_\mu)
    -\frac{\mu^2}4\,g^{\mu\alpha}_{\rm phys}
    g^{\nu\beta}_{\rm phys}\,F_{\mu\nu}F_{\alpha\beta} \right),\\
    &&F_{\mu\nu}=\partial_\mu u_\nu-\partial_\nu u_\mu.
    \end{eqnarray}
Here $\mu^2$ is the parameter having mass squared dimension and $L(g_{\mu\nu}^{\rm phys},\varphi,\partial\varphi,u_\mu)$ is a matter Lagrangian containing some direct coupling of the vector field to matter, $\partial L/\partial u_\mu\neq 0$. $F^2$-term provides a kinetic term for $u_\mu$ and guarantees absence of ghosts among the components of this vector field.

This theory is obviously Weyl invariant by the same mechanism as in \cite{ChamsMukh} and needs a conformal gauge. This gauge can be chosen in the form analogous to (\ref{gauge0}) (we consider the case of a timelike vector field)
    \begin{eqnarray}
    g^{\alpha\beta}
    u_\alpha u_\beta=-1.     \label{vectorgauge}
    \end{eqnarray}
Equations of motion in this gauge read as those of general relativity with matter sources given by ``pressureless dust fluid" of Proca vector field $u^\mu$ which has a non-uniform mass squared $m^2=\varepsilon/\mu^2$ given by the density of this fluid $\varepsilon$,
    \begin{eqnarray}
    &&G^{\mu\nu}=T^{\mu\nu}
    +\varepsilon\,u^\mu u^\nu+T_F^{\mu\nu},  \label{EinstEq}\\
    &&\mu^2\nabla_\nu F^{\nu\mu}
    -\varepsilon\,u^\mu
    +\frac{\partial L}{\partial u_\mu}=0, \label{Proca}\\
    &&\varepsilon=T+R,                    \label{density1}\\
    &&T_F^{\mu\nu}=
    \mu^2\left(F^{\mu\alpha}F^{\;\;\nu}_\alpha
    -\frac14g^{\mu\nu}F^2\right).
    \end{eqnarray}
Here the Proca field kinetic term of $u_\mu$ guarantees the absence of ghosts. Similarly to Eq.(\ref{density}) of \cite{ChamsMukh} the dust density is given by traces of the Einstein and matter stress tensors, and the conservation law for the dust fluid (obtained by differentiating (\ref{Proca})) reads
    \begin{eqnarray}
    \nabla_\mu\left(\varepsilon\,u^\mu
    -\frac{\partial L}{\partial u_\mu}\right)=0. \label{cons}
    \end{eqnarray}

In view of the generalized Proca equation (\ref{Proca}) $\varepsilon$ algebraically expresses via $\varphi$ in terms of the coupling of $u_\mu$ to matter and the rotational component $\varepsilon_{\rm rot}$,
    \begin{eqnarray}
    &&\varepsilon=
    -u_\mu\frac{\partial L}{\partial u_\mu}
    +\varepsilon_{\rm rot},       \label{1000}\\
    &&\varepsilon_{\rm rot}=
    -\mu^2u_\mu\nabla_\nu F^{\nu\mu}.
    \end{eqnarray}
The latter of course vanishes for a potential vector field with $F_{\mu\nu}=0$ and $T^{\mu\nu}_F=0$.

Now consider the effect of the potential vector field $u^\mu=\delta^\mu_0$ in the homogeneous Friedmann cosmology driven by the inflaton field $\varphi$ with the Lagrangian
    \begin{eqnarray}
    &&L(g_{\mu\nu},\varphi,\partial\varphi,u_\mu)=
    -\frac12\,(\partial_\mu\varphi)^2-V(\varphi)
    +L_{\rm int}(g_{\mu\nu},\varphi,\partial\varphi,u_\mu).
    \end{eqnarray}
It depends on the choice of the interaction between $u_\mu$ and $\varphi$. The simplest non-derivative and derivative interactions can be organized as follows
    \begin{equation}
    L_{\rm int}=
    \left\{\begin{array}{l}
    \,g^{\mu\nu}u_\mu u_\nu F(\varphi),\quad\varepsilon=2F(\varphi),\\
    \\
     \,g^{\mu\nu}u_\mu\partial_\nu\varphi\,F(\varphi)
     ,\quad\varepsilon=-u^\mu\partial_\mu\varphi\,F(\varphi)
     \end{array}\right.,
    \end{equation}
where the dust densities of the potential flow are respectively expressed according to the first term of Eq.(\ref{1000}). The first of these interactions, from the viewpoint of inflationary dynamics, is not interesting because in the gauge (\ref{vectorgauge}) it reduces to a simple modification of the inflaton potential by an extra term $F(\varphi)$. The second (derivative) type of interaction is more interesting. With the Friedmann metric $ds^2=-dt^2+a^2(t)\,d{\mathbf x}^2$ it generates the following contribution to the total matter stress tensor and dark matter density
    \begin{eqnarray}
    &&T^{\mu\nu}_{\rm int}\equiv g^{\mu\nu}L_{\rm int}+2\,\frac{\partial L_{\rm int}}{\partial g_{\mu\nu}}=F\dot\varphi\times{\rm diag}\left(1,\frac{1}{a^2}\,
    \delta^{ik}\right),\\
    &&\varepsilon=-F\dot\varphi.  \label{algebraic}
    \end{eqnarray}
In view of the relation $\varepsilon+T^{00}_{\rm int}=0$ the Friedmann equation for the Hubble parameter $H\equiv\dot a/a$ -- $00$-component of the gravitational equations (\ref{EinstEq}) -- remains unmodified by the vector field, whereas the equation for the inflaton acquires extra friction and rolling force terms (we recover the reduced Planck mass $M^2_P$ assumed to be one above),
    \begin{eqnarray}
    &&3M_P^2\,H^2=\frac12\,
    \dot\varphi^2+V, \label{00}\\
    &&\ddot\varphi+(3H+F)\,\dot\varphi+V'+3HF=0.   \label{ik}
    \end{eqnarray}
In the slow roll regime this allows one, without changing the known expression for the Hubble parameter $H\simeq\sqrt{V/3M_P^2}$, to vary the duration of inflation stage,
    \begin{eqnarray}
    \dot\varphi\simeq-\frac{V'+3HF}{3H+F},   \label{ik}
    \end{eqnarray}
by varying the magnitude and the sign of $F(\varphi)$.\footnote{Note that by adjusting the shape of $F(\varphi)$, $V'+3HF=0$, one can even get the {\em exact} de Sitter expansion with a constant $\varphi$ and $H=\sqrt{V(\varphi)/3M_P^2}$.}

However, the potential vector field with $F_{\mu\nu}=0$ cannot serve as dark matter, because its density is algebraically related to the inflaton (\ref{algebraic}) and completely decays simultaneously with the inflaton in the end of inflation (we assume that $F(\varphi)\to 0$ for $\varphi\to 0$). Yet, the role of dark matter can be played by the rotational part of the vector field which survives the decay of $\varphi$ and $\partial L/\partial u_\mu$ in (\ref{1000}). In view of (\ref{cons}) it satisfies the usual conservation law $\nabla_\mu(\varepsilon_{\rm rot}u^\mu)=0$. Under a natural assumption that the rotational part of $u^\mu$ is much smaller than its potential part $u^\mu_{(0)}\equiv\delta^\mu_0$,
    \begin{eqnarray}
    u^\mu=\delta^\mu_0+u^\mu_{\rm rot},\quad |\,u^\mu_{\rm rot}|\ll 1,
    \end{eqnarray}
this equation reduces to $\partial_0(a^3\varepsilon_{\rm rot})=0$ and gives at post-inflationary stage a typical dust evolution law $\varepsilon_{\rm rot}(t,{\bf x})=C({\bf x})/a^3(t)$ with $C({\bf x})$ accounting for inhomogeneities of the inflaton field at the end of inflation. This might simulate adiabatic perturbations of real cold dark matter.

\section{Conclusions}
Mimetic dark matter model of \cite{ChamsMukh} can be interpreted as a conformal extension of the Einstein general relativity. Gauging out the local Weyl symmetry of this theory results in an extra degree of freedom describing
a potential flow of dust which can serve as a model of real cold dark matter. The density of this dust is played by the Lagrangian multiplier for the conformal gauge in the gauged out version of the theory. It is shown that for a positive energy density of this dust the theory is free from ghost instabilities, though it can suffer from the gravitational instability associated with caustic surfaces of the geodesic flow. Positive energy criterion gives strong preference to stable configurations with a positive scalar curvature and trace of matter stress tensor, because in this theory the dark matter density is given by the sum of those.

Analogous conformal extension of the Einstein theory is suggested in the form of the generalized Proca vector field with the inhomogeneous mass parameter playing the role the dark matter density. This model includes both potential and rotational flows of the pressureless perfect fluid. Depending on the coupling of the vector field to matter, the potential part of this flow can essentially modify inflationary scenario, but cannot model modern dark matter, because its density decays simultaneously with the inflaton field in the end of inflation. The rotational part of the vector field, however, might play the role of rotating dark matter and mimic its real adiabatic perturbations. In view of optically inert nature of the real cold dark matter this model might not contradict missing observational signatures for rotating dark matter halos.

\section*{Acknowledgements}
I am deeply indebted for fruitful stimulating discussions with V.Mukhanov, V.Rubakov, S.Sibiryakov and I.Tuytin. This work was supported by the RFBR grant No. 11-02-00512.


\begin{thebibliography}{99}
\bibitem{ChamsMukh}A.H.Chamseddine and V.Mukhanov, {\em Mimetic Dark Matter}, arXiv:1308.5410.

\bibitem{FradkinVilkovisky}E.S.Fradkin and G.A.Vilkovisky, Phys. Lett. {\bf B73} (1978) 209.
    
\bibitem{Maldacena}J. Maldacena, {\em Einstein Gravity from Conformal Gravity}, arXiv:1105.5632.

\bibitem{BlaPuSib}D.Blas, O.Pujolas and S.Sibiryakov, JHEP {\bf 0910} (2009) 029, arXiv:0906.3046.

\bibitem{caustics}G.N.Felder, L.Kofman and A.Starobinsky, JHEP {\bf 0209} (2002) 026, arXiv:hep-th/0208019.

\bibitem{Mukohyama}S.Mukohyama, Phys. Rev. {\bf D80} (2009) 064005, arXiv:0905.3563.

\bibitem{Mukohyamaetal} N. Arkani-Hamed, H. C. Cheng, M. A. Luty, S. Mukohyama and T. Wiseman, JHEP {\bf 0701} (2007) 036, arXiv:hep-ph/0507120.

\end{thebibliography}
\end{document}